\documentclass[12pt,letterpaper]{article}
\usepackage[a4paper, total={7in, 10in}]{geometry}

\usepackage{graphicx}
\usepackage{helvet}
\usepackage{authblk}
\usepackage{hyperref}
\usepackage{amsmath} 
\usepackage{amssymb} 
\usepackage{orcidlink} 
\usepackage[super,comma,sort&compress]  
   {natbib}

\usepackage[utf8]{inputenc} 
\usepackage[T1]{fontenc}    
\usepackage{url}            
\usepackage{booktabs}       
\usepackage{amsfonts}       
\usepackage{nicefrac}       
\usepackage{microtype}      
\usepackage{xcolor}         
\usepackage{multirow}
\usepackage{threeparttable}
\usepackage{amsmath}
\usepackage{subcaption}
\usepackage{makecell}
\usepackage{placeins}

\usepackage{color}
\usepackage{soul}
\usepackage{xcolor}

\makeatletter
\renewcommand{\maketitle}{\bgroup\setlength{\parindent}{0pt}
\begin{flushleft}
  \textbf{\@title}
  
  \@author
\end{flushleft}\egroup}
\makeatother

\title{Simultaneous Polysomnography and Cardiotocography Reveal Temporal Correlation Between Maternal Obstructive Sleep Apnea and Fetal Hypoxia}
\date{}

\author[1,$\#$]{Jingyu Wang}
\author[2,3,$\#$,\orcidlink{0009-0008-6988-2183}]{Donglin Xie}
\author[4,$\#$]{Jingying Ma}
\author[1]{Yunliang Sun}
\author[5]{Linyan Zhang}
\author[5]{Rui Bai}
\author[5]{Zelin Tu}
\author[6]{Liyue Xu}
\author[5]{Jun Wei}
\author[5]{Jingjing Yang}
\author[6]{Yanan Liu}
\author[7]{Chuanhou Zhang}
\author[7]{Leilei Zhang}
\author[6]{Huijie Yi}
\author[6]{Bing Zhou}
\author[6]{Long Zhao}
\author[6]{Xueli Zhang}
\author[4,8]{Mengling Feng}
\author[6]{Xiaosong Dong}
\author[9]{Thomas Penzel}
\author[5,*]{Guoli Liu}
\author[2,3,10,*,\orcidlink{0000-0001-7521-5127}]{Shenda Hong}
\author[6,*]{Fang Han}

\affil[1]{Department of Respiratory and Critical Care Medicine, Shandong medical and pharmaceutical university hospital, Binzhou, China}
\affil[2]{National Institute of Health Data Science, Peking University, Beijing, China}
\affil[3]{Institute of Medical Technology, Peking University Health Science Center, Beijing, China}
\affil[4]{Saw Swee Hock School of Public Health, National University of Singapore, Singapore}
\affil[5]{Department of Obstetrics and Gynecology, Peking University People’s Hospital, Beijing, China}
\affil[6]{Division of Sleep Medicine, Peking University People's Hospital, Beijing, China}
\affil[7]{Department of Obstetrics and Gynecology, Binzhou Medical University Hospital, Binzhou, China}
\affil[8]{Institute of Data Science, National University of Singapore, Singapore}
\affil[9]{Interdisciplinary Center of Sleep Medicine, Charité - Universitätsmedizin Berlin, Berlin, Germany}
\affil[10]{Institute for Artificial Intelligence, Peking University, Beijing, China}

\affil[$\#$]{These authors contributed equally}
\affil[*]{Correspondence: guoleeliu@163.com, hongshenda@pku.edu.cn, hanfang1@hotmail.com}

\begin{document}

\maketitle

\section*{ABSTRACT}

\textbf{Background:} Obstructive sleep apnea syndrome (OSAS) during pregnancy is a prevalent complication that can negatively affect fetal outcomes. However, quantitative studies on the immediate effects of maternal hypoxia on fetal heart rate (FHR) changes are lacking, which could provide valuable clinical insights into the potential link between maternal OSAS and fetal hypoxia.

\textbf{Methods:} Time-synchronized polysomnography (PSG) and cardiotocography (CTG) data from two prospective cohorts were used to analyze the temporal correlation between maternal hypoxic events and FHR changes (accelerations or decelerations). Maternal hypoxic event characteristics, notably including hypoxic duration and the hypoxic burden area, were analyzed using generalized linear modeling (GLM) to assess their associations with different FHR alteration. To precisely capture the dynamic fetal response, a phase-specific analysis framework was employed that explicitly divided each maternal hypoxic event into pre-, during-, and post-event phases, enabling temporal mapping of FHR fluctuations in relation to maternal oxygen desaturation episodes.

\textbf{Results:} A total of 118 pregnant women were included, with 35 from Peking University People's Hospital (Center A) and 83 from the Maternal Sleep in Pregnancy and the Fetus (MSP) dataset (Center B). A chi-square test revealed a strong temporal association between maternal hypoxic events and FHR changes (all \( P < 0.0001 \)), with FHR accelerations being the predominant fetal response. Notably, a longer hypoxic duration was strongly correlated with increased FHR accelerations (\( P < 0.05 \)), while both prolonged hypoxia (Coef. \( = 0.012 \), \( P < 0.001 \)) and greater \( \text{SpO}_2 \) drop value (Coef. \( = -0.151 \), \( P = 0.046 \)) were significantly linked to FHR decelerations in the MSP dataset. Phase-specific analysis revealed a significant transient increase in mean FHR during maternal hypoxic events. In Center A, the mean FHR changed from 136.47 bpm (pre-event) to 142.28 bpm (during) and 137.29 bpm (post-event), with a similar pattern observed in Center B (135.13--140.23--133.40 bpm).

\textbf{Conclusion:} This study establishes two key findings regarding the immediate fetal response to maternal hypoxia. Firstly, a direct temporal correlation exists between maternal hypoxic events and FHR changes, predominantly compensatory accelerations, with hypoxic duration being a critical predictor of these responses. Secondly, the fetal response is dynamic and reversible, characterized by a transient but significant increase in both mean FHR and its variability during hypoxic episodes, which rapidly returns to baseline upon resolution.

\section*{KEYWORDS}


OSAS, Pregnancy, Fetal Heart Rate

\section*{INTRODUCTION}

Obstructive sleep apnea syndrome (OSAS) is a prevalent health concern during pregnancy, affecting up to 15\% of pregnant women, with severity often increasing as gestation progresses{\cite{chirakalwasan2026obstructive}}. This condition has been linked to several adverse fetal outcomes, including an increased risk of preterm birth, fetal growth restriction (FGR), and the need for cesarean delivery{\cite{maniaci2024obstructive, ansari2025obstructive, lu2021sleep, sanapo2025pregnancy, laurila2026higher}}. In addition to these immediate risks, recent studies have highlighted potential long-term health implications for the offspring of women with OSAS{\cite{morrakotkhiew2021early}}. It is essential to understand the underlying pathophysiological mechanisms contributing to its adverse effects. Characterized by recurrent episodes of partial or complete upper airway obstruction during sleep, OSAS leads to intermittent hypoxia and hypercapnia, which can significantly impact for both maternal and fetal health{\cite{ansari2025obstructive}}. Animal studies indicate that maternal exposure to these conditions can result in acute placental hypoperfusion and fetal asphyxia, further exacerbating the risks to pregnancy{\cite{ginosar2021chronic}}. Fetal heart rate (FHR) is a vital indicator of fetal oxygenation, reflecting
real-time responses to hypoxic stress. Maternal sleep apnea or hypopnea, often associated with episodes of maternal hypoxia, has been linked to significant reductions in FHR, with several case reports suggesting an association between OSAS and FGR{\cite{joel1978fetal, roush2004obstructive, sahin2008obstructive}}.

To clarify the clinical association between maternal hypoxia, fetal hypoxia, and acute FHR changes, multiple prospective studies have implemented synchronous monitoring of maternal sleep parameters and fetal cardiac activity with enlarged sample sizes. Nevertheless, conflicting results have been reported in subsequent studies{\cite{pitts2021fetal, wilson2022maternal, dipietro2023fetal, skrzypek2022fetal, fung2013effects}}. Considering the higher prevalence and greater severity of OSAS in obese pregnant women and the heightened hypoxic susceptibility of FGR fetuses, targeted validation studies have been conducted in high-risk populations. DiPietro et al.{\cite{dipietro2023fetal}} and Skrzypek et al.{\cite{skrzypek2022fetal}} replicated relevant analyses in obese pregnant women (BMI $> 30$~kg/m$^2$) and pregnancies complicated by FGR, respectively, yet no definitive evidence was found to verify that maternal hypoxic events trigger FHR decelerations in these specific cohorts.

Notably, existing studies possess limitations. First, mild and transient maternal hypoxia may trigger fetal compensatory responses, including elevated FHR and increased baseline variability, whereas few studies have explored the correlation between maternal hypoxia and FHR acceleration. Second, quantitative assessments of the area of FHR accelerations and decelerations, as well as dynamic FHR variability, are absent in current literature, which precludes an intuitive and comprehensive elucidation of the impact of maternal intermittent hypoxia on fetal cardiac rhythmicity.

To address this gap, we hypothesized that maternal sleep apnea or hypopnea, particularly when coupled with hypoxia occurrences, would have a measurable, temporal impact on FHR alterations. In this study, we utilized data from two prospective cohorts to quantitatively assess the instant effect of maternal hypoxic events on FHR, providing a more comprehensive understanding of how OSAS influences fetal health. To enrich the diversity of hypoxic event characteristics for modeling purposes, we deliberately enrolled women with suspected or confirmed OSAS alongside non-OSAS controls. OSAS patients are expected to exhibit a higher incidence and greater variability of maternal hypoxic events, which could provide valuable clinical insights into the potential link between maternal OSAS and fetal hypoxia.

\section*{METHODS}
\subsection*{Overview}
\begin{figure}[htbp]
    \centering
    \includegraphics[width=1.0\textwidth]{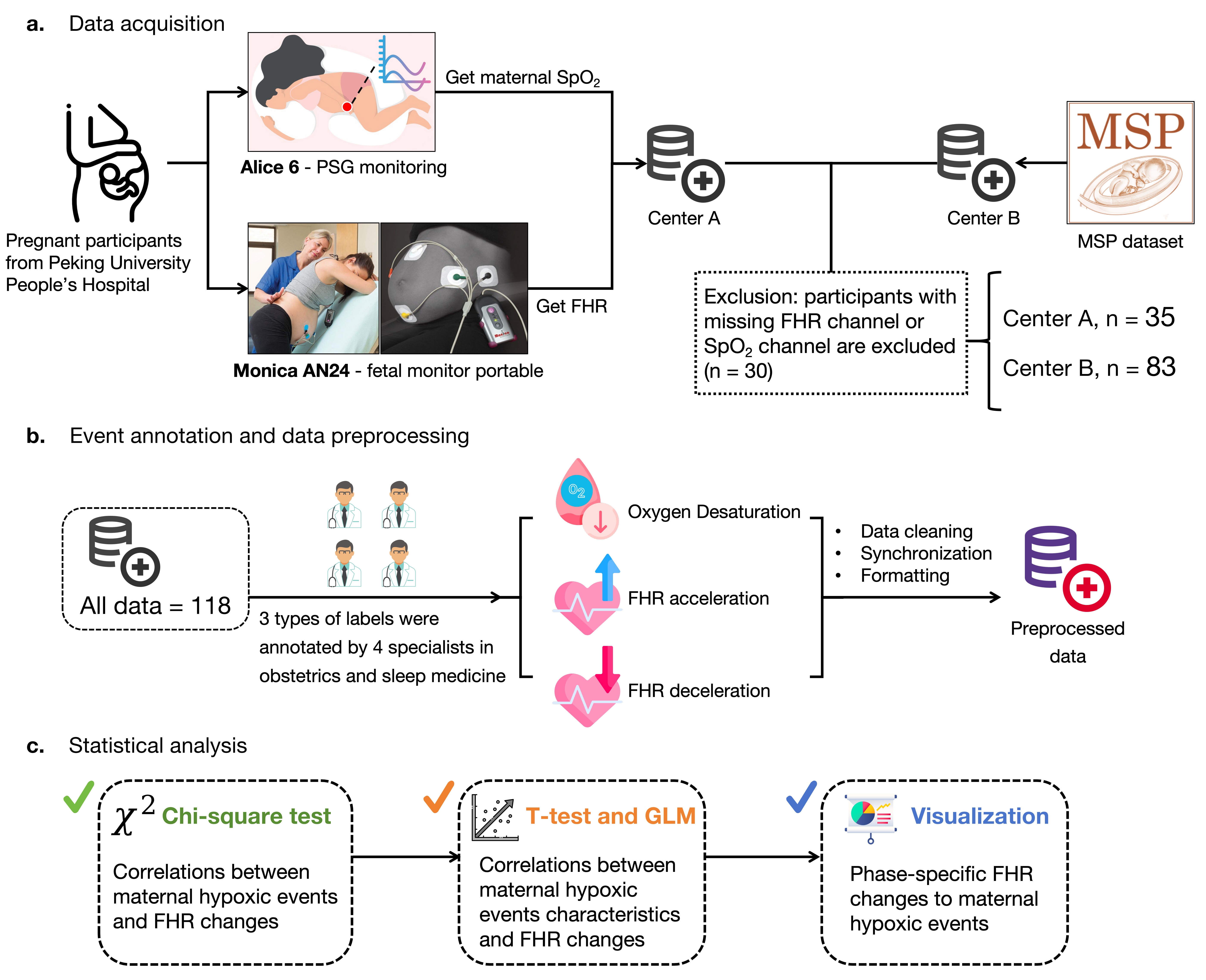}
    \caption{General overview of this study. \textbf{a}. Data from Peking University People's Hospital were collected using the same devices as the MSP dataset, with standard criteria applied for data selection. \textbf{b}. Data from both centers were professionally annotated and rigorously processed to ensure analytical accuracy. \textbf{c}. Various statistical methods were used to assess the relationship between maternal hypoxic events and FHR changes, with results visualized accordingly.}
    \label{fig:overview}
\end{figure}

The general overview of this study is illustrated in \textbf{\autoref{fig:overview}}, which comprehensively presents the procedural framework of the research. From data collection to event annotation, rigorous selection criteria were applied to ensure the representativeness of the participants and the quality of the data. In addition, the study employed thorough and methodical statistical analyses to uncover potential associations between maternal hypoxic events and FHR changes. Detailed methodologies and analyses for each part of the study are elaborated in the subsequent sections.

\subsection*{Data Acquisition and Participant Selection}
This study utilized two centers (A and B), both prospective and adhering to similar research protocols, to investigate the direct impact of OSAS on FHR responses among women in their third trimester.

\textbf{Center A}: Pregnant women reporting snoring were consecutively recruited from the obstetric clinic at Peking University People's Hospital. Inclusion criteria included: age $\geq$ 18 years, gestational age $\geq$ 28 weeks, and a singleton pregnancy. Exclusion criteria comprised complicated pregnancies with fetal anomalies, prior sleep testing or treatment for OSAS, clinically unstable medical conditions, and significant underlying pulmonary or cardiac comorbidities. This project received approval from the Institutional Review Board at Peking University People's Hospital (2021PHB416-001), and written informed consent was obtained from all participants.

\textbf{Center B}: Data were obtained from the MSP dataset. Eligibility was restricted to non-smoking women with pre-pregnancy obesity (BMI $> 30\ \text{kg/m}^2$), without previously identified sleep disorders and absent significant conditions that jeopardized the pregnant woman or fetus. Common obesity-related conditions, including pre-existing or pregnancy-induced hypertension and diabetes, were not exclusion criteria. The study was approved by the Institutional Review Board (NA\_00093511). All participants provided informed consent.

All women in both centers underwent PSG with concurrent CTG using the Monica AN24 fetal monitoring device. The monitoring duration was set to be $\geq 7$ hours. Upon completion of the monitoring, the PSG and CTG monitoring data were immediately collected and downloaded. Rigorous selection criteria were applied to the initial sample to ensure data completeness and quality. The exclusion criteria encompassed two primary aspects: first, participants with missing maternal oxygen saturation or FHR signal channels were excluded; second, samples with prolonged periods of monitoring failure or zero signal recordings, or with significant abnormal fluctuations in maternal oxygen saturation or FHR signal channels, were also excluded. Following this screening process, a total of 35 complete cases from center A and 83 complete cases from center B were included in the final analysis.

\subsection*{Data Collection and Event Annotation}
\subsubsection*{Demographic information}
Basic demographic information were collected, including maternal and gestational age, BMI, systolic and diastolic blood pressure and medical comorbidities such as gestational diabetes.

\subsubsection*{Polysomnography}
Overnight PSG for Datasets A was performed in the sleep laboratory at Peking University 
People’s Hospital, utilizing the Compumedics E series (Abbotsford, Victoria, Australia) 
and the Alice 6 system (Philips Respironics, Inc., United States). In Dataset B, all signals 
were continuously recorded using the RemLogic 1.3 N7000 data acquisition system 
(Natus Medical Inc., Broomfield, Colorado, United States).

PSG was performed according to the recommendations of the American Academy of 
Sleep Medicine. The following signals were recorded: electroencephalogram 
(F3M2, F4M1, C3M2, C4M1, O1M2, O2M1), bilateral electrooculogram, chin muscle 
electromyogram, rib cage and abdominal movement, electrocardiogram (lead 1), snoring, 
body position, bilateral anteriotibialis electromyograms, and heart rate and oxygen 
saturation by pulse oximetry. Airflow was monitored with a nasal pressure cannula and 
oronasal thermistor.

PSG was manually scored using the 2012 scoring criteria from the American Academy of 
Sleep Medicine\cite{berry2012rules} by experienced sleep technologists who were blinded to the 
fetal monitoring results. Apneas were defined as $\geq 90\%$ reduction in airflow from baseline 
for at least 10 seconds on the oronasal thermistor signal. Obstructive apneas were 
characterized by the presence of respiratory effort during the events, while central apneas 
were identified by the absence of respiratory effort. Mixed apneas were defined as those 
where respiratory effort was initially absent but reappeared in the latter part of the episode. Hypopnea events were defined with $\geq 30\%$ reduction in airflow from baseline for 
$\geq 10$ seconds associated with $\geq 3\%$ reduction in oxygen saturation and/or an arousal. 
Apnea-hypopnea index (AHI) on PSG was calculated as the average number of apneas and 
hypopneas per hour of sleep.
OSAS was defined as an AHI of $\geq 5$ events per hour, consistent with the American Academy of Sleep Medicine (AASM) 2012 guidelines{\cite{berry2012aasm}}, which represent the widely accepted clinical threshold for diagnosing OSAS in adults, including pregnant women.

\subsubsection*{FHR Monitoring}
CTG was performed to measure the continuous FHR using the Monica AN24 (Monica Healthcare Ltd., Nottingham, UK) and was time-synchronised to the PSG recording. In Center A, CTG was conducted continuously over a 10-hour period (20:00 -- 06:00), whereas monitoring duration parameters in Center B were unspecified in the protocol. This device has been validated against intrapartum monitoring utilizing scalp electrodes and with antepartum CTG\cite{dipietro2022measuring, tamber2020systematic}. The Monica AN24 is a non-invasive device that uses five adhesive electrodes placed on the maternal abdomen to track fetal electrocardiogram (ECG), maternal ECG, and uterine activity. The results of CTGs were deidentified and evaluated by two obstetricians, blinded to the sleep study results. They precisely annotated for the onset and offset of accelerations and decelerations. In cases where discrepancies in annotation occurred, a senior physician conducted a review. Definitions for FHR tracing based on guidelines from the NICHD and ACOG\cite{arnold2017intrapartum}.

\subsection*{The definition of the temporal correlation between maternal and fetal events}
\begin{figure}[htbp]
    \centering
    \subfloat{\textbf{(a)} %
        \includegraphics[width=0.8\textwidth]{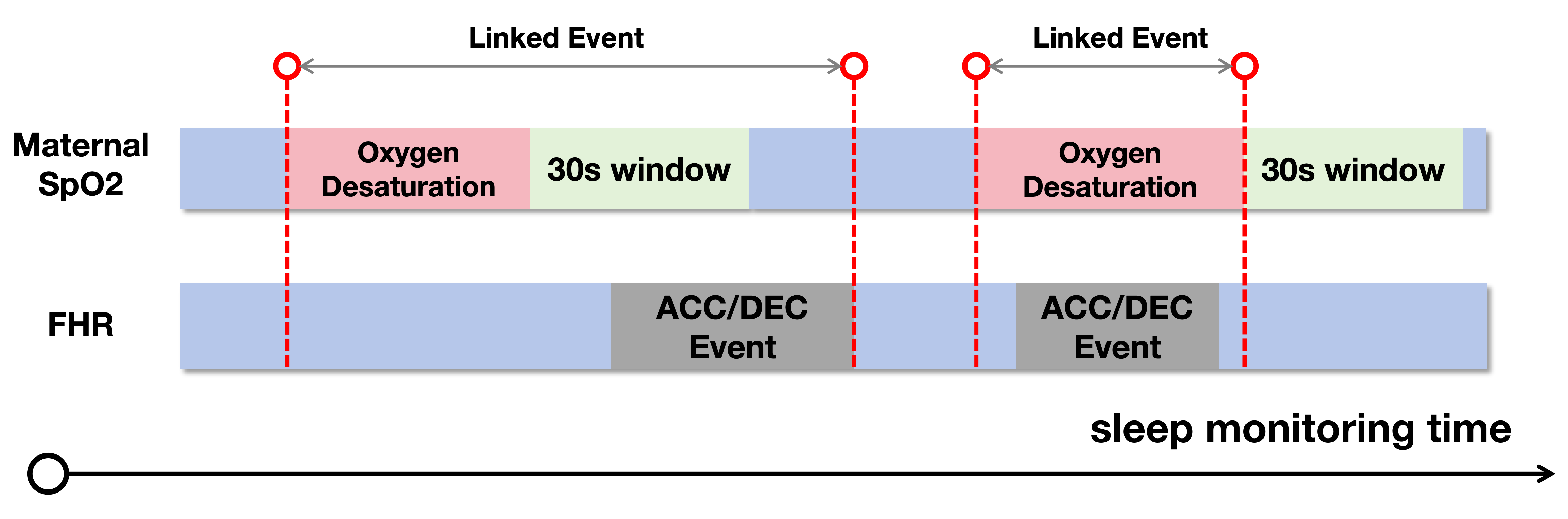}%
    }\\
    
    \subfloat{\textbf{(b)} %
        \includegraphics[width=0.8\textwidth]{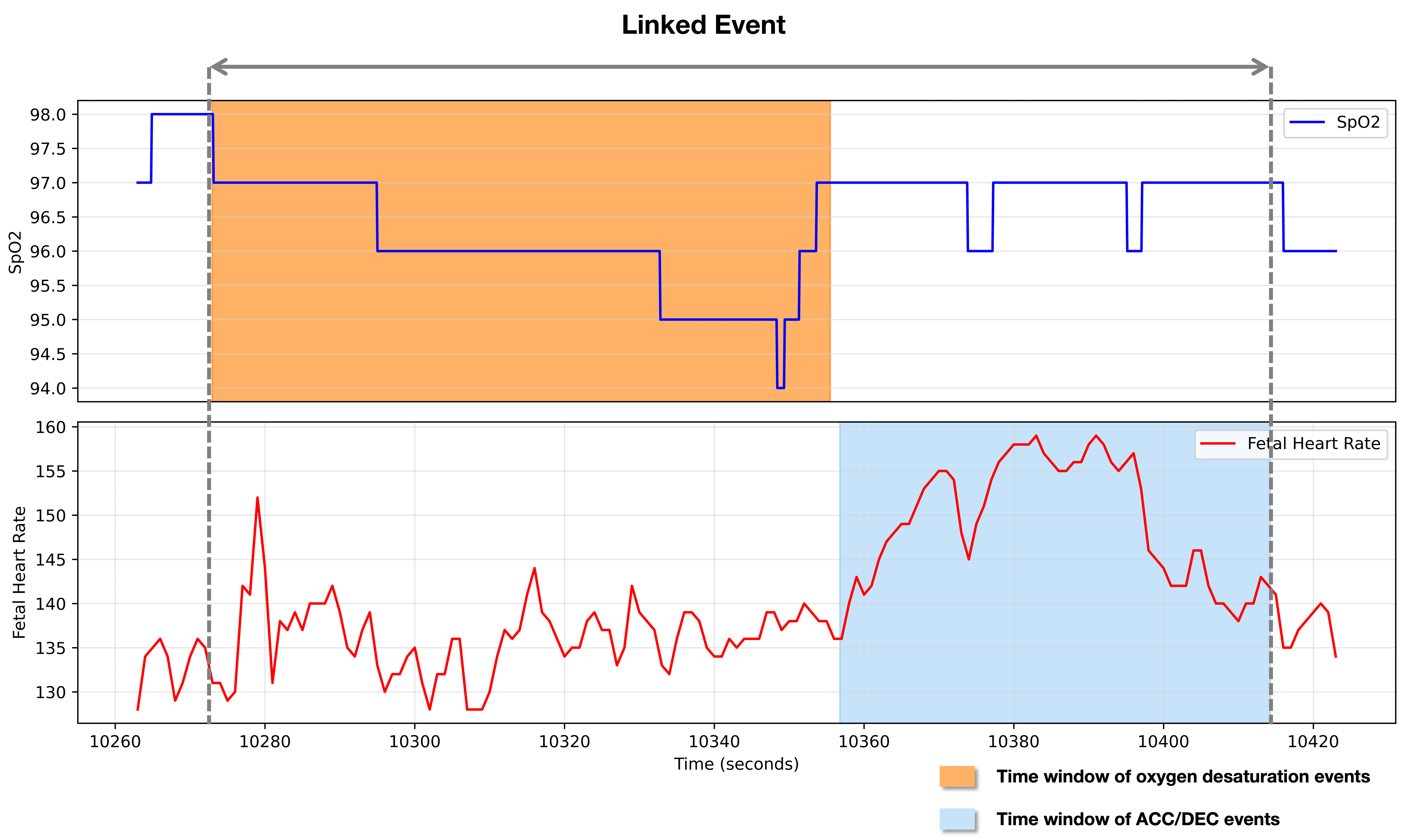}%
    }\\
    
    \caption{Actual waveform characteristics of a linked event.
    \textbf{(a)} Illustration of the linked events. 
    \textbf{(b)} Waveform characteristics of a linked event.}
    \label{fig:fhr}
\end{figure}

Pitts et al\cite{pitts2021fetal} found that the majority of both late and prolonged FHR decelerations occurred within 30 seconds of maternal respiratory events. The temporal relationship between maternal and fetal events (linked events) was defined as the occurrence of a fetal deceleration within 30 seconds following a maternal respiratory event. If multiple respiratory events occurred within the 30-second period preceding the fetal deceleration, all were considered. \textbf{\autoref{fig:fhr}} illustrates this relationship in \textbf{\autoref{fig:fhr}.a} and presents a specific case demonstrating the actual waveform characteristics of a linked event in \textbf{\autoref{fig:fhr}.b}.

As illustrated in \textbf{\autoref{fig:fhr}}, a linked event is defined as a FHR acceleration or deceleration (ACC/DEC) event that occurs within a 30-second time window following the end of a maternal hypoxic event. This time window accounts for the potential delay in the fetal response to maternal hypoxia. If another maternal hypoxic event occurs within this 30-second window, the events are merged into a continuous maternal hypoxic event. This approach ensures a comprehensive capture of the temporal relationship between maternal hypoxic events and FHR changes.

The duration of a linked event is mathematically defined as follows:
\begin{equation}
T_{\text{Linked Event}} = \max \left( T_{\text{ACC/DEC, End}} - T_{\text{MHE, Start}},\ 
T_{\text{MHE, End}} - T_{\text{MHE, Start}} \right)
\end{equation}
where {$T_{\text{ACC/DEC, End}}$} represents the termination time of the FHR acceleration or deceleration event, {$T_{\text{MHE, Start}}$} indicates the onset time of the maternal hypoxic event, and {$T_{\text{MHE, End}}$} denotes the termination time of the maternal hypoxic event.

Intuitively, this formula computes the total temporal span from the onset 
of the maternal hypoxic event ({$T_{\text{MHE, Start}}$}) to whichever endpoint 
occurs later: the termination of the maternal hypoxic event itself 
({$T_{\text{MHE, End}}$}) or the termination of the associated FHR response 
({$T_{\text{ACC/DEC, End}}$}). This accounts for two possible scenarios, 
as illustrated in \mbox{\textbf{\autoref{fig:fhr}a}}: 
(i) if the FHR acceleration or deceleration event extends beyond the end of 
the maternal hypoxic event (i.e., \mbox{$T_{\text{ACC/DEC, End}} > T_{\text{MHE, End}}$}), 
the linked event duration is determined by the FHR response offset; 
(ii) if the FHR response terminates before or coincides with the end of 
the maternal hypoxic event, the linked event duration equals the hypoxic 
event duration. The formula can equivalently be expressed as:
\begin{equation*}
T_{\text{Linked Event}} = \max\!\left(T_{\text{ACC/DEC, End}},\; T_{\text{MHE, End}}\right) 
- T_{\text{MHE, Start}}
\end{equation*}
which makes explicit that the linked event always begins at {$T_{\text{MHE, Start}}$} 
and ends at the later of the two offsets.

\subsection*{Statistical Analysis}
\subsubsection*{Chi-square test}
In this study, a chi-square test was conducted to investigate the relationship between maternal hypoxic events and FHR acceleration or deceleration events. We recorded the start and end times of each hypoxic event, as well as the corresponding times of FHR acceleration and deceleration events, and constructed a 2×2 contingency table for each case based on the overlap between these events. To analyze the relationship between maternal hypoxic events and FHR acceleration or deceleration events, a 2×2 contingency table was constructed with the following components:
\begin{itemize}
\item A1: The total duration of linked events. For the specific definition, refer to "The definition of the temporal correlation between maternal and fetal events" mentioned above.
\item A2: The total duration of maternal hypoxic events without accompanying FHR acceleration or deceleration events.
\item B1: The total duration of FHR acceleration or deceleration events without accompanying maternal hypoxic events.
\item B2: The total duration in which neither maternal hypoxic events nor FHR acceleration or deceleration events occur, representing time segments free of any correlated or independent events.
\end{itemize}

After calculating the 2×2 contingency tables for all individual cases, we further aggregated each individual's contingency table to form an overall 2×2 contingency table. It is important to note that individual cases lacking data for FHR acceleration/deceleration events were not included in the aggregated 2×2 contingency table to prevent biases from invalid data. Subsequently, the aggregated contingency table was subjected to a chi-square test to assess whether there was a statistically significant relationship between maternal hypoxic events and FHR acceleration or deceleration. By analyzing the chi-square test results, including the chi-square value and p-value, we determined whether a significant association exists between maternal hypoxic events and the occurrence of FHR acceleration or deceleration. If the p-value is less than 0.05, it is considered that there is a significant association between hypoxic events and FHR acceleration or deceleration events. The chi-square test is calculated using the following formula:
\begin{equation}
\chi^2 = \sum \frac{(O_i - E_i)^2}{E_i}
\end{equation}
where $O_i$ represents the observed frequency, and $E_i$ represents the expected 
frequency. Specifically, in this study, the formula is expressed as:
\begin{equation}
\chi^2 = \frac{(A1 - E_{A1})^2}{E_{A1}} + \frac{(A2 - E_{A2})^2}{E_{A2}} 
+ \frac{(B1 - E_{B1})^2}{E_{B1}} + \frac{(B2 - E_{B2})^2}{E_{B2}}
\end{equation}
The expected frequency for each cell is calculated under the assumption of independence between maternal hypoxic events and FHR changes: \mbox{$E_{A1} = \frac{(A1+A2)(A1+B1)}{N}$}, and similarly for the others, where $N$ is the total sum of durations.

\subsubsection*{T-test}
In statistical analysis, an independent samples t-test was conducted to compare the maternal hypoxic characteristics between the Peking University People's Hospital dataset and the MSP dataset. Four key characteristics were analyzed: $\text{SpO}_2$ nadir (event-level measures), $\text{SpO}_2$ drop value, duration of the hypoxic event (Duration), and hypoxic burden area.

\begin{figure}[htbp]
    \centering
    \includegraphics[width=0.8\textwidth]{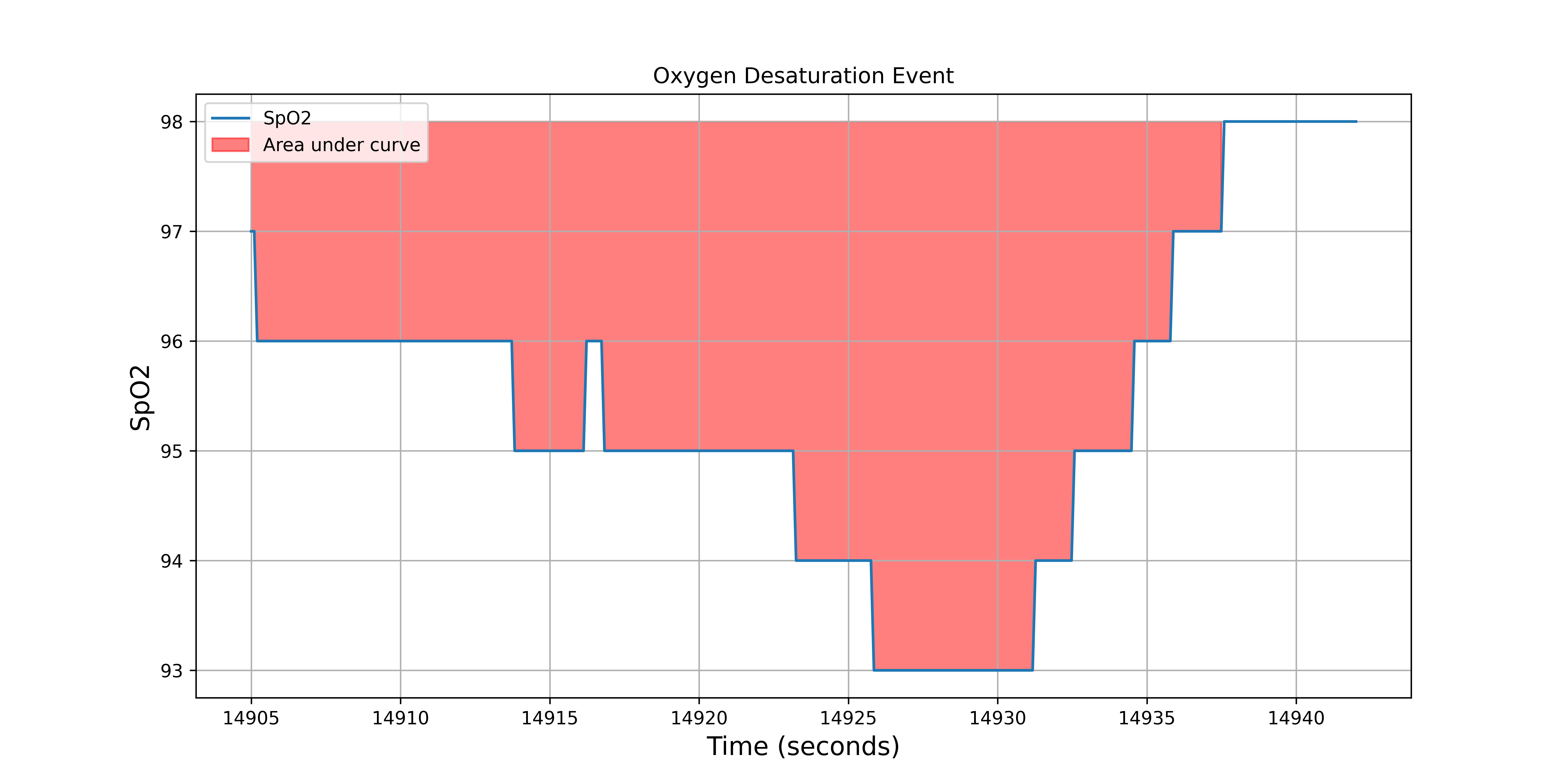}
    \caption{Schematic diagram of hypoxic burden area}
    \label{fig:hypoxic_burden}
\end{figure}

Specifically, the hypoxic burden area quantifies the cumulative oxygen desaturation during each hypoxic event by measuring the area under the $\text{SpO}_2$ curve where the values fall below a defined baseline. The baseline $\text{SpO}_2$ level ($S_{\text{baseline}}$) was defined as the maximum $\text{SpO}_2$ value recorded during the hypoxic event window itself, i.e., between the event's onset time ($t_{\text{start}}$) and end time ($t_{\text{end}}$). This reflects the initial oxygenation level at the start of the event, allowing for standardized estimation of the desaturation burden across varying events. The hypoxic burden area for a given episode is then computed as:

\begin{equation}
\text{Hypoxic Burden Area} = \int_{t_{\text{start}}}^{t_{\text{end}}} (S_{\text{baseline}} - S(t)) \, dt
\end{equation}

where $S(t)$ denotes the instantaneous $\text{SpO}_2$ level during the hypoxic event. The area is calculated using Simpson's numerical integration method and expressed in ``$\% \cdot \text{s}$", quantifying the magnitude and duration of desaturation relative to the event's baseline. This integrated metric reflects the overall physiological impact of maternal oxygen desaturation. \textbf{\autoref{fig:hypoxic_burden}} illustrates the concept.

\subsubsection*{Generalized Linear Model (GLM)}
A generalized linear model (GLM) was employed to further explore the relationships 
between maternal oxygen hypoxic event characteristics and the occurrence of FHR 
linked events, FHR acceleration events, and FHR deceleration events. Specifically, 
four key hypoxic event characteristics—hypoxic duration, SpO$_2$ nadir, SpO$_2$ 
drop value, and hypoxic burden area—were treated as independent variables, while 
the occurrence of FHR linked events, FHR acceleration events, and FHR deceleration 
events were treated as binary outcome variables. Separate univariate GLM models were 
constructed to evaluate the association between each characteristic and the three 
types of FHR changes.

The model inputs included the values of each maternal hypoxic event characteristic 
and the corresponding binary outcomes of FHR change types (1 indicating the 
occurrence of the event, 0 indicating no occurrence). The GLM utilized a logit 
link function, with the model represented as:
\begin{equation}
\log \left( \frac{p}{1 - p} \right) = \beta_0 + \beta_1 X
\end{equation}
where $p$ represents the probability of the target FHR event occurring, $X$ 
denotes the value of a specific maternal hypoxic event characteristic, and 
$\beta_0$ and $\beta_1$ are regression coefficients.

The model fitting and parameter estimation were implemented using Python's 
statsmodels library. The output included regression coefficients (Coef.), 
standard errors (Std.Err.), z-values, and p-values. Separate univariate regression 
analyses were performed for the three types of FHR changes, quantifying the impact 
of maternal hypoxic event characteristics on fetal physiological responses.

\subsubsection*{Visualization Analysis}
To further investigate the impact of maternal hypoxic events on FHR variations, box plots were used for visualization analysis. For each maternal hypoxic event, the event was divided into three phases: the 10-second period before the event onset (pre-event), the entire duration of the event (during-event), and the 10-second period following the event (post-event). Based on these time windows, the mean, standard deviation (STD), and coefficient of variation (CV) of FHR were calculated for each phase.

Specifically, for each hypoxic event, the mean, standard deviation, and coefficient of variation of FHR were calculated for the three phases (pre-event, during-event, and post-event) separately in both datasets (Center A and Center B). The calculated statistics for the mean, standard deviation, and coefficient of variation were subsequently visualized using box plots to illustrate the distribution of FHR across the different phases. This approach provided a clear visualization of the immediate effects of maternal hypoxic on FHR, particularly with regard to fluctuations and stability in FHR patterns.

\section*{RESULTS}
\subsection*{Clinical characteristics of participants}
This study included a total of 118 cases, with 35 participants from Center A (Peking University People's Hospital) and 83 from Center B (the Maternal Sleep in Pregnancy and Fetus [MSP] dataset)\cite{dipietro2023fetal}\cite{zhang2018national}\cite{dipietro2021fetal}.

In Center A, a total of 42 pregnant women were initially enrolled. One participant withdrew due to intolerance to PSG monitoring, and 6 cases were excluded for invalid FHR data, leaving 35 for final analysis. Based on PSG results (\(\text{AHI} \geq 5\) events/h), the participants were divided into two groups: 17 women (48.6\%) in the OSAS group and 18 women (51.4\%) in the non-OSAS group. The OSAS group exhibited significantly higher BMI and AHI values compared to the non-OSAS group (both \( P < 0.01 \)).

In Center B, 106 pregnant women with valid PSG data were enrolled from MSP dataset\cite{dipietro2023fetal}\cite{zhang2018national}\cite{dipietro2021fetal}. After excluding 23 participants due to invalid FHR data, 83 women were included in the final analysis. According to PSG results, 38 women (45.8\%) were classified into the OSAS group, while 45 women (53.6\%) were in the non-OSAS group. The OSAS group demonstrated significantly higher BMI, AHI, and REM latency compared to the non-OSAS group (all P < 0.01). Clinical and PSG results from both groups are summarized in \textbf{\autoref{tab:clinical_psg}}.

\begin{table}[htbp]
    \centering
    \small
    \renewcommand{\arraystretch}{1.5}
    \caption{Clinical and PSG results from two centers. \textbf{PSG}: polysomnography; \textbf{OSAS}: obstructive sleep apnea syndrome; \textbf{BMI}: body mass index; \textbf{BP}: blood pressure; \textbf{AHI}: apnea-hypopnea index; \textbf{REM}: rapid eye movement sleep; \textbf{TST}: total sleep time. \textit{a} indicates a statistically significant difference when OSAS group compared to the non-OSAS group in center A. \textit{b} indicates a statistically significant difference when OSAS group compared to the non-OSAS group in center B.}
    \label{tab:clinical_psg}
    \begin{tabular}{lcccc}
        \toprule
        \textbf{Groups} & \multicolumn{2}{c}{\textbf{Center A}} & \multicolumn{2}{c}{\textbf{Center B}} \\
        \cmidrule(r){2-3} \cmidrule(r){4-5}
         & OSAS (n=17) & Non-OSAS (n=18) & OSAS (n=38) & Non-OSAS (n=45) \\
        \midrule
        Age, year & 36.29 ± 5.65 & 34.11 ± 3.12 & 28.95 ± 6.67 & 25.04 ± 5.44 \\
        Gestational age, week & 33.76 ± 5.38 & 31.06 ± 6.53 & 36.25 ± 0.79 & 36.34 ± 0.81 \\
        BMI, kg/m$^2$ & 32.01 ± 5.33\tnote{a} & 26.92 ± 2.82 & 42.53 ± 7.59\tnote{b} & 40.50 ± 4.10 \\
        Systolic BP, mmHg & 120.5 ± 12.7 & 119.7 ± 13.2 & 115.30 ± 9.79 & 113.87 ± 8.30 \\
        Diastolic BP, mmHg & 75.5 ± 10.5 & 77.1 ± 9.4 & 70.70 ± 8.83 & 69.71 ± 7.75 \\
        Hypertension & 14 (82.4\%) & 10 (55.6\%) & 5 (11.11\%) & 10 (26.31\%) \\
        Diabetes & 10 (58.8\%) & 4 (22.2\%) & 3 (6.67\%) & 6 (15.79\%) \\
        AHI, events/h & 12.77 ± 7.51\tnote{a} & 2.36 ± 1.24 & 14.54 ± 11.74\tnote{b} & 1.93 ± 1.42 \\
        Mean SpO$_2$, \% & 95.41 ± 1.28 & 95.94 ± 1.16 & 96.61 ± 1.35 & 96.66 ± 1.02 \\
        Sleep efficiency, \% & 73.16 ± 12.24 & 75.98 ± 15.50 & 74.25 ± 15.94 & 71.27 ± 13.09 \\
        REM Latency, min & 138.13 ± 66.29 & 135.88 ± 85.81 & 121.62 ± 83.63\tnote{b} & 93.66 ± 56.11 \\
        Total Sleep Time, min & 341.60 ± 61.24 & 356.98 ± 73.77 & 358.79 ± 80.39 & 337.58 ± 59.02 \\
        3 Stage/TST, \% & 21.08 ± 7.65 & 19.28 ± 6.92 & 17.67 ± 10.27 & 18.12 ± 8.74 \\
        REM Sleep/TST, \% & 14.38 ± 6.51 & 16.82 ± 3.71 & 15.48 ± 5.91 & 14.59 ± 6.19 \\
        \bottomrule
    \end{tabular}
\end{table}

\subsection*{Correlations between maternal hypoxic events and FHR changes}
In Center A, a total of 425 linked events (the definition is detailed in Fig. 5) were recorded during the monitoring of 35 pregnant women, of which 381 were acceleration-related events and 44 were deceleration-related events. Notably, 86.4\% (38/44) of the events linked to deceleration occurred in one woman with severe OSAS (AHI = 31.4 events/hour, ODI = 34.3 events/hour, lowest SpO$_2$ = 79\%, and 8 minutes 27 seconds of oxygen saturation (SpO$_2$) below 90\%). In Center B, a total of 313 linked events (235 were acceleration-linked events, and 78 were deceleration-linked events) were recorded. Among the 78 deceleration-linked events, 19 individuals were involved, with 14 women classified as non-OSAS and 5 individuals diagnosed with OSAS. Furthermore, 3 pregnant women with the highest number (N=9,8) of deceleration-linked events were all non-OSAS individuals.

A chi-square test was conducted to assess the relationship between maternal hypoxic events and FHR acceleration or deceleration events, with the detailed results presented in \textbf{\autoref{tab:chi_square_results}}. Through synthesizing the results from centers A and B, the chi-square test indicates a statistically significant association between maternal hypoxic events and FHR changes (linked events). Across all centers, the linked events exhibit high chi-square values and extremely low p-values (all \textit{P}<0.0001), suggesting that this association is not attributable to random variation and is statistically significant. 

\begin{table}[htbp]
    \centering
    \small
    \caption{Chi-square test results of different fypes of linked events}
    \label{tab:chi_square_results}
    \begin{threeparttable}
    \renewcommand{\arraystretch}{1.5} 
    \begin{tabular}{llcccccc}
        \toprule
        \textbf{Center} & \textbf{Linked Event Type} & \textbf{A1 (s)} & \textbf{A2 (s)} & \textbf{B1 (s)} & \textbf{B2 (s)} & \textbf{$\chi^2$} & \textbf{\textit{P}-value} \\
        \midrule
        A & Deceleration & 2497.75 & 29532.3 & 7742.75 & 692466.2 & >10.83 & <0.001 \\
        A & Acceleration & 25977.85 & 16865.7 & 164394.75 & 525000.7 & >10.83 & <0.001 \\
        B & Deceleration & 2915.7 & 86543.1 & 15213.25 & 2026977.95 & >10.83 & <0.001 \\
        B & Acceleration & 17139.6 & 78186 & 134134.25 & 1902190.15 & >10.83 & <0.001 \\
        \bottomrule
    \end{tabular}
        \end{threeparttable}
\end{table}

The analysis was based on 2×2 contingency tables from both centers, with the detailed chi-square test results presented in \textbf{\autoref{tab:chi_square_results}}. The variables in the contingency table were defined as follows: A1: The total duration of linked events; A2: The total duration of maternal hypoxic events without accompanying FHR acceleration or deceleration events; B1: The total duration of FHR acceleration or deceleration events without accompanying maternal hypoxic events; B2: The total duration during which neither maternal hypoxic events nor FHR acceleration or deceleration events occur, representing time segments free of any correlated or independent events.

\subsection*{Correlations between maternal hypoxic events characteristics and FHR changes}
Descriptive statistics for hypoxic event characteristics in center A and B are presented in 
\textbf{\autoref{tab:hypoxic_stats}}. The analysis shows that significant differences 
(92.43 $\pm$ 2.22 vs. 93.04 $\pm$ 2.61) were observed in SpO\textsubscript{2} nadir (event-level measures), 
duration of the hypoxic event (37.35 $\pm$ 26.37 vs. 31.06 $\pm$ 21.03), and hypoxic burden area 
(93.61 $\pm$ 85.13 vs. 79.22 $\pm$ 94.65) between the two centers (\textit{p} < 0.05), while no significant difference 
was observed in the SpO\textsubscript{2} drop value (4.10 $\pm$ 1.83 vs. 4.11 $\pm$ 2.16).

\begin{table}[htbp]
    \centering
    \small
    \caption{Descriptive statistics for maternal hypoxic event characteristics}
    \label{tab:hypoxic_stats}
    \renewcommand{\arraystretch}{1.5} 
    \begin{tabular}{lcccc}
        \toprule
        & \textbf{Center A} & \textbf{Center B} & \textbf{t-Statistics} & \textbf{\textit{P} Value} \\
        \midrule
        Participants, n & 35 & 83 & & \\
        Number of hypoxic events, n & 1425 & 3252 & & \\
        SpO\textsubscript{2} nadir, \% & 92.43 $\pm$ 2.22 & 93.04 $\pm$ 2.61 & -8.07 & <0.0001 \\
        SpO\textsubscript{2} Drop Value, \% & 4.10 $\pm$ 1.83 & 4.11 $\pm$ 2.16 & -0.18 & 0.86 \\
        Hypoxic Duration, s & 37.35 $\pm$ 26.37 & 31.06 $\pm$ 21.03 & 7.86 & <0.0001 \\
        Hypoxic Burden Area & 93.61 $\pm$ 85.13 & 79.22 $\pm$ 94.65 & 3.44 & 0.0006 \\
        \bottomrule
    \end{tabular}
\end{table}

To further investigate the potential associations between the characteristics of maternal 
hypoxic events and FHR changes, this study analyzed four key hypoxic event 
characteristics—hypoxic duration, SpO\textsubscript{2} nadir (event-level measures), 
SpO\textsubscript{2} drop value, and hypoxic burden area— and their relationship with 
the occurrence of FHR-linked events, FHR acceleration events, and FHR deceleration events. 
A generalized linear model (GLM) was employed to evaluate the statistical correlation of 
each characteristic with these three types of FHR changes, quantifying the impact of maternal 
hypoxia on fetal physiological responses. The results are summarized in 
\textbf{\autoref{tab:glm_results}}, which reports the regression coefficients (Coef.), standard errors (Std.Err.), and {\textit{P}}-values for each maternal hypoxic characteristic across the two centers, 
illustrating the statistical relevance in explaining different FHR change patterns.

\begin{table}[htbp]
    \centering
    \renewcommand{\arraystretch}{1.5} 
    \caption{GLM analysis results for maternal hypoxic event characteristics and FHR changes. \textbf{Coef. (Coefficient)}: The coefficient represents the direction and magnitude of the effect that the independent variable has on the dependent variable. \textbf{Std. Err. (Standard Error)}: The standard error of the estimated coefficient, reflecting the uncertainty of the coefficient estimation. A smaller standard error indicates a more stable and reliable estimate. \textbf{\textit{P} value}: The \textit{P} value used to test the statistical significance of the estimated coefficient. A \textit{P} value less than 0.05 suggests that the result is statistically significant.}
    \label{tab:glm_results}
    \begin{tabular}{llcccccc}
        \toprule
        \multicolumn{2}{l}{\textbf{Center}} & \multicolumn{3}{c}{\textbf{A}} & \multicolumn{3}{c}{\textbf{B}} \\
        \cmidrule(lr){3-5} \cmidrule(lr){6-8}
        \textbf{Statistic} & & \textbf{Coef.} & \textbf{Std.Err.} & \textbf{\textit{P} value} & \textbf{Coef.} & \textbf{Std.Err.} & \textbf{\textit{P} Value} \\
        \midrule
        \multicolumn{8}{c}{\textbf{Maternal hypoxic events related to FHR changes}} \\
        \midrule
        Hypoxic Duration & & 0.014 & 0.002 & 0.000 & 0.008 & 0.002 & 0.000 \\
        SpO\textsubscript{2} nadir & & 0.002 & 0.026 & 0.952 & 0.019 & 0.011 & 0.081 \\
        SpO\textsubscript{2} Drop Value & & 0.031 & 0.030 & 0.311 & -0.083 & 0.028 & 0.003 \\
        Hypoxic Burden Area & & 0.003 & 0.001 & 0.000 & 0.000 & 0.000 & 0.426 \\
        \midrule

        \multicolumn{8}{c}{\textbf{Maternal hypoxic events related to FHR acceleration events}} \\
        \midrule
        Hypoxic Duration & & 0.014 & 0.002 & 0.000 & 0.006 & 0.002 & 0.019 \\
        SpO\textsubscript{2} nadir & & 0.007 & 0.026 & 0.773 & 0.011 & 0.011 & 0.299 \\
        SpO\textsubscript{2} Drop Value & & 0.017 & 0.031 & 0.583 & -0.065 & 0.029 & 0.024 \\
        Hypoxic Burden Area & & 0.003 & 0.001 & 0.000 & 0.000 & 0.000 & 0.959 \\
        \midrule

        \multicolumn{8}{c}{\textbf{Maternal hypoxic events related to FHR deceleration events}} \\
        \midrule
        Hypoxic Duration & & -0.005 & 0.010 & 0.663 & 0.012 & 0.003 & 0.000 \\
        SpO\textsubscript{2} nadir & & 0.064 & 0.118 & 0.588 & 0.023 & 0.012 & 0.053 \\
        SpO\textsubscript{2} Drop Value & & 0.130 & 0.095 & 0.174 & -0.151 & 0.075 & 0.046 \\
        Hypoxic Burden Area & & -0.001 & 0.003 & 0.771 & 0.001 & 0.000 & 0.138 \\
        \bottomrule
    \end{tabular}
\end{table}

In the regression analysis for FHR-linked events, Center A demonstrated significant 
statistical associations for the duration of maternal hypoxic events 
(Coef. = $0.014$, $P < 0.001$) and hypoxic burden area (Coef. = $0.003$, $P < 0.001$), 
while in Center B, the duration of hypoxic events remained significant 
(Coef. = $0.008$, $P < 0.001$). For FHR acceleration-linked events, significant effects of hypoxic duration were 
observed in both centers (Center A: Coef. = $0.014$, $P < 0.001$; Center B: 
Coef. = $0.006$, $P = 0.019$). However, for FHR deceleration-linked events, the MSP center revealed statistical 
significance for hypoxic duration (Coef. = $0.012$, $P < 0.001$) and 
SpO\textsubscript{2} drop value (Coef. = $-0.151$, $P = 0.046$), while no significant 
associations were found at Center A. The GLM results provide crucial insights into the mechanisms through which maternal 
hypoxic events influence FHR patterns, laying a foundation for further exploration of 
the complex dynamics in maternal-fetal oxygenation.

\subsection*{Phase-specific FHR changes to maternal hypoxic events}

\begin{figure}[htbp]
    \centering
    \includegraphics[width=\textwidth]{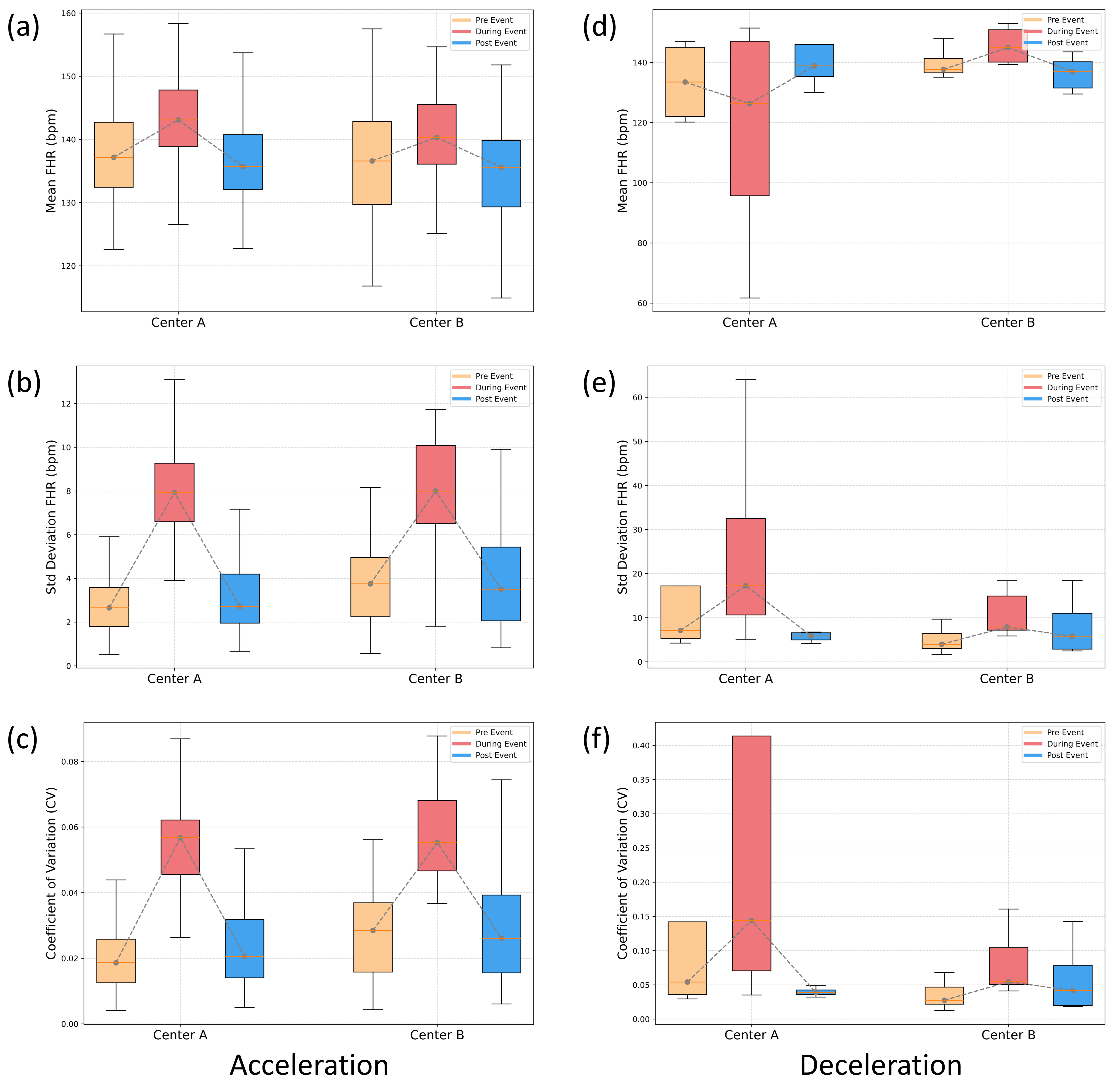}
    \caption{Phase-specific FHR metrics stratified by response type. 
    Left column (a--c): acceleration-linked events; right column (d--f): deceleration-linked events. 
    \textbf{(a, d)} Mean FHR across pre-, during-, and post-event phases in Center A and Center B. 
    \textbf{(b, e)} FHR standard deviation (FHR\textsubscript{SD}) across the three phases. 
    \textbf{(c, f)} Coefficient of variation (CV) across the three phases. 
    Each box represents the interquartile range; the dashed line connects within-group medians 
    across phases to highlight temporal trends.}
    \label{fig:FHR_comparison}
\end{figure}

Box plots were used to visualize FHR changes before, during, and after maternal hypoxic events, 
stratified by FHR response type (acceleration vs. deceleration). Each maternal hypoxic event was 
divided into three phases: the 10-second perihypoxic period before the event onset (pre-event), 
the entire duration of the hypoxic event (during-event), and the 10-second perihypoxic period 
following the event offset (post-event). For each hypoxic event, the mean FHR, standard deviation 
(FHR\mbox{\textsubscript{SD}}), and coefficient of variation (CV) were calculated across these three phases. 
The resulting distributions are displayed in \mbox{\textbf{\autoref{fig:FHR_comparison}}} and summarized in 
\mbox{\textbf{\autoref{tab:fhr_phase_summary}}}.

\begin{table}[htbp]
\centering
\renewcommand{\arraystretch}{1.5}
\caption{Summary of phase-specific FHR metrics for acceleration- and deceleration-linked events 
in both centers. Values represent means across all events.}
\label{tab:fhr_phase_summary}
\small
\begin{tabular}{llcccc}
\hline
\textbf{Event Type} & \textbf{Metric} & \textbf{Center} & \textbf{Pre-event} & \textbf{During-event} & \textbf{Post-event} \\
\hline
\multirow{6}{*}{Acceleration}
 & \multirow{2}{*}{Mean FHR (bpm)}
   & A & 136.65 & 143.84 & 136.98 \\
 & & B & 135.41 & 139.95 & 133.44 \\
\cline{2-6}
 & \multirow{2}{*}{FHR\textsubscript{SD} (bpm)}
   & A & 3.87  & 9.19  & 5.00  \\
 & & B & 5.13  & 10.93 & 6.37  \\
\cline{2-6}
 & \multirow{2}{*}{CV}
   & A & 0.054 & 0.065 & 0.038 \\
 & & B & 0.051 & 0.081 & 0.056 \\
\hline
\multirow{6}{*}{Deceleration}
 & \multirow{2}{*}{Mean FHR (bpm)}
   & A & 133.54 & 116.43 & 142.38 \\
 & & B & 133.24 & 142.09 & 133.17 \\
\cline{2-6}
 & \multirow{2}{*}{FHR\textsubscript{SD} (bpm)}
   & A & 15.35 & 25.90 & 5.67  \\
 & & B & 10.49 & 11.40 & 14.03 \\
\cline{2-6}
 & \multirow{2}{*}{CV}
   & A & 0.124 & 0.340 & 0.040 \\
 & & B & 0.115 & 0.083 & 0.146 \\
\hline
\end{tabular}
\end{table}


\paragraph{Acceleration-linked events.}
For FHR acceleration-linked events, a consistent and symmetric temporal pattern was observed 
in both centers (\mbox{\textbf{\autoref{fig:FHR_comparison}a}}). Mean FHR rose from 136.65~bpm (pre-event) 
to 143.84~bpm (during-event) and returned to 136.98~bpm (post-event) in Center~A, 
and from 135.41~bpm to 139.95~bpm and back to 133.44~bpm in Center~B. 
Correspondingly, FHR\mbox{textsubscript{SD}} increased from 3.87 to 9.19~bpm (Center~A) and from 
5.13 to 10.93~bpm (Center~B) during the event, with rapid recovery to near-baseline levels 
within 10~seconds of event resolution (\mbox{\textbf{\autoref{fig:FHR_comparison}b}}). 
The CV followed a similar transient elevation pattern in both centers (\mbox{\textbf{\autoref{fig:FHR_comparison}c}}). 
These results indicate that maternal hypoxic events are associated with a transient, reversible 
increase in both mean FHR and variability when the dominant fetal response is acceleration, 
consistent with sympathetically mediated compensatory cardiac output augmentation.

\paragraph{Deceleration-linked events.}
The phase-specific profiles differed notably between the two centers for deceleration-linked events (\mbox{\textbf{\autoref{fig:FHR_comparison}d--f}}), likely reflecting their distinct patient compositions.

In \mbox{\textbf{Center~A}}, mean FHR decreased from 133.54~bpm (pre-event) to 116.43~bpm 
(during-event), followed by a post-event rebound to 142.38~bpm, which exceeded the pre-event 
baseline. This pattern is physiologically consistent with a deceleration occurring \mbox{\emph{during}} 
the maternal hypoxic event itself, followed by a compensatory rebound upon resolution. 
However, the markedly elevated FHR\mbox{\textsubscript{SD}} during the event (25.90~bpm vs. 
15.35~bpm pre-event) and the high CV (0.340) indicate substantial inter-event variability. 
This is attributable to the small sample size in Center~A: of the 44 deceleration-linked events, 
86.4\% originated from a single participant with severe OSAS (AHI = 31.4~events/h, lowest 
SpO\mbox{\textsubscript{2}} = 79\%), limiting the generalizability of this sub-group estimate.

In \mbox{\textbf{Center~B}}, mean FHR showed a modest increase during the maternal hypoxic event 
(133.24 to 142.09~bpm), returning to pre-event levels post-event (133.17~bpm). 
This pattern suggests that, in deceleration-linked events observed in Center~B, the fetal 
response during the maternal hypoxic episode is predominantly an initial compensatory 
acceleration, while the deceleration manifests within the 30-second post-event window---the 
temporal lag that defines a ``linked event.'' Such a two-phase fetal response---early 
sympathetically driven acceleration followed by a subsequent deceleration---is consistent 
with the brainstem-mediated chemoreceptor cascade in response to progressive 
hypoxemia{\cite{jia2023pathophysiological}}. FHR\mbox{\textsubscript{SD}} and CV in Center~B 
showed comparatively smaller fluctuations across phases, reflecting the greater sample 
diversity (78 deceleration events across 19 participants).

To assess whether the CV metric provides information beyond a simple 
scaling of FHR\mbox{\textsubscript{SD}} by mean FHR, we fitted a linear regression 
model (\mbox{$\sigma = \alpha\mu + \beta_{\text{phase}} + \gamma$}) using phase as 
a categorical predictor and tested the joint significance of the phase terms 
via an F-test after controlling for mean FHR. For acceleration-linked events, 
the phase term was highly significant in both Center~A 
(\mbox{$p = 1.10 \times 10^{-9}$}) and Center~B (\mbox{$p = 1.16 \times 10^{-8}$}), 
indicating that FHR\mbox{\textsubscript{SD}} depends on event phase independently 
of mean FHR level. The partial correlation between phase and 
FHR\mbox{\textsubscript{SD}} (controlling for mean FHR) was not significant 
in either center (Center~A: \mbox{$r = 0.068$}, \mbox{$p = 0.343$}; 
Center~B: \mbox{$r = 0.014$}, \mbox{$p = 0.841$}), consistent with the 
non-monotonic (inverted-U) shape of the phase effect, which is 
not captured by a linear correlation coefficient. For deceleration-linked 
events, the phase term was not significant in either center 
(Center~A: \mbox{$p = 0.802$}; Center~B: \mbox{$p = 0.197$}), suggesting that 
CV in the deceleration subgroup approximates a scaled version of 
FHR\mbox{\textsubscript{SD}} rather than providing additional phase-specific information.

Taken together, the phase-specific analyses demonstrate that maternal hypoxic events are 
temporally associated with significant and reversible alterations in FHR dynamics, with the 
nature of these alterations depending on the dominant fetal response type and the severity 
of the underlying hypoxic burden.

\section*{DISCUSSION}
OSAS during pregnancy, characterized by intermittent hypoxia, has been associated with various maternal and fetal complications. This study aims to provide clinical evidence of the vertical effects of maternal intermittent hypoxia on the fetus by synchronously acquiring PSG and CTG in late-pregnancy women. We find that there are correlations between maternal hypoxic events and FHR changes, with the changes primarily characterized by accelerations. The duration of maternal hypoxic events is significant factor influencing immediate FHR changes. The second finding relates to the continuous monitoring of FHR and associated variables before, during, and after maternal hypoxic events. Both the mean FHR and its variability are significantly elevated during hypoxic episodes, but rapidly return to baseline levels following the resolution of oxygen desaturation. The hypothesis test results further clarify the informational content of 
the CV metric. For acceleration-linked events, the significant phase 
effect on FHR\mbox{\textsubscript{SD}} after controlling for mean FHR confirms 
that FHR variability is not merely a proportional reflection of the 
baseline heart rate level, but exhibits a distinct phase-dependent 
elevation during maternal hypoxic events. This implies that the during-event 
increase in FHR variability reflects a genuine physiological response---likely 
driven by heightened autonomic activity---rather than a mathematical artefact 
of elevated mean FHR. In contrast, for deceleration-linked events, the 
absence of a significant phase effect suggests that FHR variability changes 
in proportion to mean FHR changes, and the CV does not provide additional 
discriminative information in this subgroup. Given the smaller and more 
heterogeneous sample of deceleration events, particularly in Center~A, 
this null result should be interpreted with caution.

Several research groups have conducted prospective studies using PSG or portable sleep monitoring devices in conjunction with FHR monitoring in small sample sizes to investigate whether maternal hypoxia can induce fetal hypoxia and immediate changes in FHR. Pitts et al.~\cite{pitts2021fetal} performed simultaneous monitoring with the WatchPAT portable sleep monitor and the Monica AN24 in 40 late-pregnancy women, 18 of whom (9 with OSAS) experienced 37 late deceleration events, 84\% of which were linked to respiratory events. These preliminary data suggest a strong correlation between maternal respiratory events and fetal decelerations. In contrast, Fung et al.~\cite{fung2013effects} found no significant association between respiratory events and FHR decelerations, except for one case of delayed deceleration in a fetus with severe FGR. Similarly, Wilson et al.~\cite{wilson2022maternal} found no correlation between respiratory events and FHR changes.

Additionally, OSAS is more prevalent and severe in obese pregnant women, whose fetuses with FGR may be more susceptible to hypoxia. DiPietro and Skrzypek's team replicated these studies in specific populations of obese women (BMI $>30$ kg/m\textsuperscript{2}) and those with FGR, but found no conclusive evidence linking respiratory events to fetal 
decelerations in these populations~\cite{dipietro2023fetal, skrzypek2022fetal}. These conflicting findings highlight the need for further research. Firstly, mild and transient hypoxia may trigger compensatory changes, potentially increasing fetal FHR baseline and variability. However, none of these studies have examined the relationship between maternal hypoxic events and FHR acceleration. Secondly, the correlation between detailed maternal hypoxic parameters and FHR has not been quantitatively analyzed.

In this study, the patient populations differed between the two centers. The majority of 
women in Center B are African American, while all women in Center A are Chinese, with a 
lower BMI and earlier gestational age. Using quantitative analyses across these two distinct 
populations, the study found consistent results, indicating maternal hypoxic events are associated 
with alterations in FHR, typically manifesting as accelerations. 
The divergent GLM results for deceleration events between the two centers 
may reflect differences in sample size and population characteristics. 
Center A comprised 35 participants with 44 deceleration-linked events, 
of which 86.4\% were concentrated in a single participant with severe OSAS 
(AHI = 31.4 events/h, lowest SpO\mbox{\textsubscript{2}} = 79\%), substantially 
limiting statistical power for this subgroup. Center B, with 83 participants 
and 78 deceleration-linked events distributed across 19 individuals, provided 
greater event diversity and statistical power to detect the significant 
associations between hypoxic duration, SpO\mbox{\textsubscript{2}} drop value, 
and FHR deceleration observed in the GLM analysis. 
This finding underscores the importance of adequate sample size and 
event heterogeneity when modeling less frequent FHR responses such as decelerations.
The placenta acts as a "respiratory organ," enabling the exchange of fetal blood gases 
with the environment via passive oxygen transfer from maternal blood, driven by a concentration 
gradient\cite{valverde2024effects}. Maternal hypoxia, especially when it is prolonged or severe, plays a critical 
role in influencing immediate FHR changes. Given that fetal oxygen reserves are approximately $42$ mL and consumption is around $20$ mL/min, the fetal oxygen supply can only sustain for about 
2 minutes in a state of complete oxygen deprivation\cite{rurak2013changes}. Therefore, fetal oxygenation is 
highly dependent on adequate maternal oxygen levels. 
Maternal hypoxia, caused by respiratory events, can lead to fetal hypoxia, triggering a 
cascade of physiological responses that ultimately result in changes to FHR. 

In cases of mild, short-duration maternal hypoxia, the fetus can typically compensate by increasing cardiac output through catecholamine-mediated heart rate elevations and redistribution of oxygenated blood to vital organs, thereby maintaining aerobic metabolism\cite{khalyfa2017late}. This is facilitated by fetal hemoglobin, which, due to its superior oxygen affinity 
compared to adult hemoglobin, enables more efficient oxygen transport and utilization even under 
low oxygen partial pressures. The concentration of fetal hemoglobin (approximately $180$--$220$ g/L) 
is significantly higher than that of adults ($110$--$140$ g/L), further supporting the fetus' ability 
to endure transient hypoxic stress\cite{jia2023pathophysiological}. 
In these situations, transient mild hypoxia may activate chemoreceptors in the carotid 
body and aortic arch, leading to reflexive increases in perfusion to essential fetal organs 
such as the brain, heart, adrenal glands, and placenta. This is often reflected as transient 
heart rate accelerations due to sympathetic stimulation mediated by the brainstem, resulting 
in increased baseline and variability of FHR.

However, when hypoxia is prolonged or recurrent, the fetus may no longer sustain 
adequate perfusion to vital organs through peripheral vasoconstriction and 
centralization of blood flow\cite{jia2023pathophysiological}. In such cases, the brainstem-mediated chemoreceptor reflexes can induce heart rate deceleration, likely as a mechanism to reduce myocardial workload. 
Additionally, humoral factors, such as vasoconstrictors released by the adrenal 
glands in response to hypoxemia, can further modulate norepinephrine release, 
influencing FHR dynamics. As hypoxia progresses, fetal bradycardia and decreased baseline variability may occur, 
potentially leading to metabolic acidosis. These findings suggest that while the 
fetus can typically tolerate mild and transient hypoxia, prolonged or severe hypoxic 
conditions may overwhelm compensatory mechanisms, resulting in decompensatory 
deceleration and impaired fetal well-being.

Notably, even in pregnant women without a formal diagnosis of OSAS, correlations between FHR changes and maternal intermittent hypoxia were observed over time. This suggests that intermittent hypoxia, stemming from maternal respiratory events, can occur even in the absence of clinically diagnosed OSAS. The protective effects of progesterone on central respiratory control may play a role in mitigating the impact of such events. Importantly, decelerations were frequently noted in the non-OSAS group, with the majority of these decelerations temporally linked to maternal respiratory disturbances. Although these women did not meet the diagnostic threshold for OSAS, many exhibited mild respiratory disruptions, indicating that even subtle maternal respiratory disturbances may adversely affect fetal well-being. Current diagnostic criteria for OSAS during pregnancy, primarily based on the AHI derived from the general population, do not adequately consider the duration and severity of respiratory events, nor the magnitude and duration of hypoxia. These limitations suggest that AHI alone may not fully capture the severity of OSAS or its maternal-fetal consequences in pregnancy. The findings highlight the need for a reconsideration of the AHI thresholds applied to pregnant women, suggesting that lower thresholds may be clinically more relevant for this population. This underscores the importance of developing new diagnostic indices and standards specifically tailored to pregnant women.

This study reflects an enhanced responsiveness of FHR to maternal hypoxic events and suggests that maternal hypoxic may induce instability in FHR, particularly in the during-event phase. Furthermore, the observed tendency of mean FHR and fluctuations to stabilize in the post-event phase indicates a return to a relatively steady physiological state for the fetus after the maternal hypoxic event ends. The combined analyses from the chi-square test and box plot visualizations support the conclusion that maternal hypoxic events significantly influence FHR acceleration or deceleration events. Specifically, characteristics of maternal hypoxic events, such as duration, significantly impact FHR variation patterns, manifested as an increase in mean FHR and enhanced fluctuations during the maternal hypoxia period. These findings provide valuable insights into the complex effects of maternal hypoxic events on FHR and offer critical information for potential clinical interventions.

There are multiple strengths to our study. Firstly, this study lies in the integration of laboratory-based PSG with simultaneous CTG during overnight to investigate the temporal relationship between maternal hypoxic events and FHR changes. Secondly, we utilized data from two centers, which included diverse ethnic populations, to ensure broader generalisability. Furthermore, we conducted a comprehensive and objective quantitative analysis of maternal events in relation to time-synchronized CTG, enabling an in-depth evaluation of the linked events. This approach indicates a potential cause-and-effect relationship, though it cannot establish causality, and highlights a likely contributory association.

The interpretation of our study should take into account its limitations. Intermittent fetal tracings or data loss resulted in several maternal hypoxic events without corresponding fetal data for correlation. While the absence of fetal data may be considered a limitation, it does not affect the overall interpretation of our results, as the primary focus was on the temporal relationship between maternal and fetal events, rather than the total number of events. Besides, the sample size of this study was constrained by the inherent challenges of conducting overnight PSG in pregnant populations—a limitation exacerbated in late gestation due to maternal discomfort, frequent nocturnal awakenings and physiological changes. Furthermore, the requirement for concurrent continuous CTG intensified physical and psychological burdens, contributing to a 50\% attrition rate in recruitment. Moreover, our temporal association analysis relies on predefined time windows rather than more rigorous signal coupling methods such as cross-correlation analysis, Granger causality testing, or transfer entropy. While the chosen approach is clinically interpretable and aligned with prior literature, future studies employing more noise-robust time-series coupling techniques would strengthen causal inference.

\section*{CONCLUSION}
This prospective cohort study recruiting two distinct maternal populations demonstrated temporal correlations between maternal hypoxic episodes and acute changes in FHR. Specifically, mild and transient maternal hypoxia primarily triggered compensatory FHR accelerations and enhanced beat-to-beat variability, and these fetal cardiac alterations could rapidly reverse to baseline levels once maternal oxygen saturation returned to normal. In contrast, prolonged duration or severe severity of maternal hypoxic events was identified as an independent risk factor for pathological FHR decelerations. Notably, subtle yet measurable fetal cardiac reactivity to intermittent maternal hypoxia was also observed in pregnant women without OSAS, who were stratified as normal cases based on conventional AHI cut-off values. Collectively, these findings solidly evidence that standalone AHI measurement fails to achieve accurate stratification of fetal hypoxia risk among pregnant individuals. Based on synchronous PSG-CTG recordings acquired under real clinical scenarios, the present study further elaborates the acute dynamic coupling mechanism of oxygen transport between mothers and fetuses in patients with gestational OSAS. Further controlled interventional studies are warranted to clarify whether targeted therapeutic interventions for gestational OSAS can stabilize fetal cardiovascular function and mitigate hypoxia-related fetal adverse responses.

\section*{ACKNOWLEDGEMENTS}
Data provided for this report was funded by National Institutes of Health grant R01 HD079411, awarded to Janet A. DiPietro, Johns Hopkins Bloomberg School of Public Health, Johns Hopkins University. The National Sleep Research Resource was supported by the U.S. National Institutes of Health, National Heart Lung and Blood Institute (R24 HL114473, 75N92019R002).

Funding: This work was supported by the National Natural Science Foundation of China (No.62102008) to SDH, (No. 82500145) to JYW and National Natural Science Foundation of China Regional Innovation and Development Joint Fund (No.U20A20388) to GLL.

\newpage

\bibliography{reference}

\bigskip


\newpage

\end{document}